\documentclass[prb,aps,amssymb,twocolumn,showpacs,floatfix]{revtex4}
\usepackage[dvips]{graphicx}
\def\be{\begin{equation}}
\def\ee{\end{equation}}
\def\bea{\begin{eqnarray}}
\def\eea{\end{eqnarray}}
\def\ve{\varepsilon}
\def\bp{{\bf p}}

\begin{document}

\title{Oscillatory  ac- and photoconductivity
of a 2D electron gas: Quasiclassical transport beyond the Boltzmann equation}
\author{I.A.~Dmitriev$^{1,*}$}
\author{A.D.~Mirlin$^{1,2,\dagger}$}
\author{D.G.~Polyakov$^{1,*}$}
\affiliation{$^{1}$Institut f\"ur Nanotechnologie,
Forschungszentrum Karlsruhe, 76021 Karlsruhe, Germany \\
$^{2}$Institut f\"ur Theorie der Kondensierten Materie, Universit\"at
Karlsruhe, 76128 Karlsruhe, Germany}

\begin{abstract}

We have analyzed the quasiclassical mechanism of magnetooscillations
in the ac- and photoconductivity, related to non-Markovian dynamics of
disorder-induced electron scattering.  While the magnetooscillations
in the photoconductivity are found to be weak, the effect manifests
itself much more strongly in the ac conductivity, where it may easily
dominate over the oscillations due to the Landau quantization.  We
argue that the damping of the oscillatory photoconductivity provides a
reliable method of measuring the homogeneous broadening of Landau
levels (single-particle scattering rate) in high-mobility structures.

\end{abstract}


\maketitle

\section{Introduction}
\label{sec1}

An intriguing development in the study of a high-mobility
two-dimensional electron gas (2DEG) was the recent observation
\cite{zudov01} of magnetooscillations of the photoconductivity of a
sample subjected to microwave radiation, as a function of the ratio
$\omega/\omega_c$. Here $\omega$ and $\omega_c$ are the radiation
frequency and the cyclotron frequency, respectively. Subsequent
experiments, \cite{mani02,zudov03,dorozhkin03,willett03} working with
very-high-mobility samples, yielded yet another dramatic discovery:
for sufficiently high radiation power, the minima of the oscillations
evolve into ``zero-resistance states", i.e., the dissipative
resistance of a sample becomes vanishingly small.

The nature of the oscillations in the photoconductivity $\sigma_{\rm
ph}$ has raised a lot of interest. An important step was made in
Ref.~\onlinecite{andreev03}, where a direct connection between the emergence
of the zero-resistance states and the oscillations was
emphasized. Specifically, it was recognized that {\it whatever} the
nature of the oscillations, when they become so large that the linear
{\it dc} response theory yields a negative $\sigma_{\rm ph}$, an
instability is developed leading to the formation of domains of
counter-flowing currents and thus to the zero measured resistance.
Following this approach, a key issue which needs to be settled for
understanding the experiments is the microscopic mechanism of the
oscillatory photoconductivity (OPC).

Similarly to the conventional Shubnikov--de Haas oscillations, the
growing body of theoretical work is focused on the oscillations of the
density of states (DOS) induced by the Landau quantization as an
essential element of the construction.  The mechanism of the OPC
identified in Ref.~\onlinecite{dmitriev03} and analyzed in detail in
Ref.~\onlinecite{dmitriev03a} hinges on the oscillations of the DOS and is
related to a radiation-induced change of the electron distribution
function in energy space, $f(\varepsilon)$, such that $f(\varepsilon)$
oscillates with varying both $\varepsilon/\omega_c$ and
$\omega/\omega_c$.  A hallmark of this contribution to $\sigma_{\rm
ph}$ is that it yields an amplitude of the OPC which is proportional
to the inelastic relaxation time $\tau_{\rm ee}$ due to
electron-electron collisions (more effective at low temperatures than
electron-phonon scattering).  Another mechanism of the OPC, based on
the effect of radiation on impurity scattering in the presence of the
Landau quantization, was suggested in Ref.~\onlinecite{durst03} (an earlier,
closely related variant of this approach was formulated in
Ref.~\onlinecite{ryzhii}).  A systematical theory of this contribution to
$\sigma_{\rm ph}$ was constructed in Ref.~\onlinecite{vavilov03}. Comparing
the results of Refs.~\onlinecite{dmitriev03,dmitriev03a} and
Ref.~\onlinecite{vavilov03}, one sees that the mechanism
\cite{dmitriev03,dmitriev03a} dominates, i.e., leads to much stronger
oscillations, if $\tau_{\rm ee}\gg\tau_q$, where $\tau_q$ is the
single-particle relaxation time due to impurity scattering.  For
typical experimental parameters, a characteristic ratio $\tau_{\rm
ee}/\tau_q\sim 10^2$. \cite{dmitriev03a}  Overall the results of
Ref.~\onlinecite{dmitriev03a} are in good agreement with the experimental
data as regards the behavior of $\sigma_{\rm ph}$ in the range of
parameters where the OPC is not too strongly damped, i.e., where the
experimental efforts have been focused so far.  In particular,
Ref.~\onlinecite{dmitriev03a} explains the emergence of strong oscillations
and, in combination with Ref.~\onlinecite{andreev03}, the formation of
zero-resistance regions.

While the agreement between theory and experiment is very encouraging, the
situation is not so clear in the experimental limit of {\it weak} (strongly
damped) oscillations. Central to the identification of the microscopic
mechanism of the oscillations is, on top of their period and phase, the
behavior of their envelope with decreasing magnetic field $B$. For any
mechanism based on the DOS oscillations, the relation between the DOS and OPC
damping factors is critically important. The OPC
\cite{dmitriev03,dmitriev03a,vavilov03} is damped at low $B$ by a factor $\exp
({-\pi/\omega_c\tau_{\rm ph}})$, where the ratio $\tau_{\rm ph}/\tau_q=1/2$ is
a distinctive feature not sensitive to microscopic details of either disorder
or weak inelastic interactions. \cite{sdh} However, as emphasized in
Ref.~\onlinecite{dmitriev03}, the experimentally reported values of $\tau_q$
and $\tau_{\rm ph}$ do not satisfy this relation, with $\tau_{\rm ph}$
noticeably larger than $\tau_q/2$, roughly by a factor of 10 in
Ref.~\onlinecite{zudov03} and by a factor of 3 in
Ref.~\onlinecite{mani02}. Taken at face value, the difference would mean that
the amplitude of the OPC observed at small $B$ is orders of magnitude higher
than given by the mechanism based on the Landau quantization, which might be
considered as a hint about a different origin of the OPC at small
$B$. Alternatively, the experiments on the damping of Shubnikov-de Haas
oscillations might overestimate the single-particle scattering rate
$\tau_q^{-1}$, e.g., because of inhomogeneous (due to macroscopic
inhomogeneities) broadening of Landau levels. To resolve this dilemma, it is
desirable to examine a range of mechanisms of the OPC in the absence of the
DOS oscillations.

In this paper, we analyze a mechanism of the OPC governed by {\it
quasiclassical} memory effects. These are related to non-Markovian
correlations in electron dynamics. \cite{mem} We assume that
$\pi/\omega_c\tau_q\gg 1$ and completely neglect weak oscillations of the
DOS. The OPC induced by the memory effects is not specific to any particular
type of disorder; however, below we concentrate on the following two-component
model, \cite{mirlin01} where the memory effects are particularly prominent. We
assume that there is a smooth random potential of remote donors that are
separated by a large spacer $d\gg k_F^{-1}$, where $k_F$ is the Fermi
wavevector, from the 2DEG plane and, in addition, there are rare short-range
scatterers, e.g., residual impurities located at or near the interface. We
consider the case $\tau_S\ll \tau_L$, where $\tau_S$ and $\tau_L$ are the
zero-$B$ momentum relaxation times due to the short-range scatterers and the
long-range disorder, respectively. From the experimental point of view, this
choice is motivated by reports (see, e.g., Ref.~\onlinecite{umansky}) that the
zero-$B$ mobility in very-high-mobility structures is frequently
limited by residual impurities and $\tau_L/\tau_S$ can be as large as 10.
Although $\tau_S\ll\tau_L$ in our model, we assume that $\tau_q$ is determined
by the smooth disorder, i.e., $\tau_q\simeq \tau_L/(2k_Fd)^2\ll \tau_S$.

The paper is organized as follows. First, in Sec.~\ref{sec2}, we outline the
approach to the photoconductivity based on the Boltzmann equation. In
Sec.~\ref{sec3}, we discuss the mechanism of the photoconductivity related to
electron-electron interactions.  In Sec.~\ref{sec4}, we turn to the
magnetooscillations induced by the memory effects. Our central results are
presented in Secs.~\ref{sec5},\ref{sec6}. Section \ref{sec5} deals with the
oscillations in the {\it ac} conductivity.  Finally, in Sec.~\ref{sec6}, we
compare two mechanisms of the oscillatory photoconductivity, quasiclassical
and quantum, related to the memory effects and the Landau quantization,
respectively.

\section{Photoconductivity: Essentials}
\label{sec2}

A necessary input to the calculation of the {\it quasiclassical} OPC
is the memory effects, discarded in the Boltzmann equation. However,
to set up a systematic formalism, it is instructive to begin with a
derivation of $\sigma_{\rm ph}$ within the conventional kinetic
theory.  The Boltzmann equation for the distribution function
$g(p,\phi,t)$ of electrons in momentum space reads:
\be
Lg(p,\phi,t)\,=\,-{\bf F}\,\partial_\bp\,g(p,\phi,t)~, 
\label{1} 
\ee
where $L\,=\,\partial_t +\omega_c\,\partial_\phi-I_{\rm el}-I_{\rm
in}$, ${\bf F}=-e(\vec{\cal E}_{\rm dc} + \vec{\cal E}_\omega
\cos\omega t)$, $\vec {\cal E}_{\rm dc}$ is the {\it dc} electric
field, $\vec{\cal E}_\omega$ is the {\it ac} field, $\phi$ is the
angle of the momentum $\bp$ with respect to the direction of
$\vec{\cal E}_{\rm dc}$, $I_{\rm el}$ and $I_{\rm in}$ are the elastic
and inelastic collision integrals, respectively.

We expand the distribution function at energy $\ve$ in a series:
$g(p,\phi,t)=\sum_{\nu n} g_{\nu n}(\ve)\exp(i \nu\phi + i n\omega
t)$. Elastic collisions lead to relaxation of angular harmonics with
$\nu\ne 0$; in particular, $I_{\rm el}g_{1 n}=-\tau^{-1}g_{1 n}$,
where $\tau$ is the momentum relaxation time. Inelastic
electron-electron collisions tend to equilibrate electrons among
themselves but are not capable of establishing a steady-state dc
photoconductivity. For the quasiclassical OPC (in contrast to that
based on the DOS oscillations, cf.\ Ref.~\onlinecite{dmitriev03a}),
the inelastic transitions due to electron-electron interaction do not
play any essential role and will be neglected. To dissipate energy
absorbed from the {\it ac} field, we introduce coupling to a thermal
bath, e.g., to an equilibrium phonon system, characterized by a
relaxation time $\tau_{\rm in}$. Under the assumption that both
$\tau_{\rm in}$ and the momentum relaxation time due to the coupling
to the bath are much longer than $\tau$, the main role of the
inelastic scattering is to yield a slow relaxation of the isotropic
($\nu=0$) part of $g$ to the equilibrium Fermi distribution $f_F$ at a
bath temperature $T$.

\begin{figure}
\centerline{ 
\includegraphics[width=8cm]{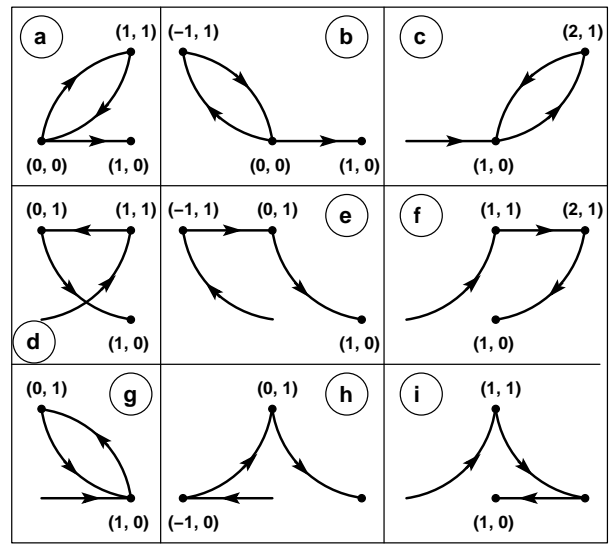}}
\vspace{3mm}
\caption{ 
Graphic representation  
of various contributions to the photoconductivity
[Eqs.~(\ref{g}),(\ref{j})] in the $(\nu,n)$ space 
at the lowest order $\sigma_{\rm
ph}\sim{\cal O}({\cal E}_{\rm dc}^0{\cal E}_\omega^2)$.} 
\label{f1}
\end{figure}

Expanding the nonequilibrium distribution function in powers of the
driving force, we have
\be
g=\sum\nolimits_{m=0}(-L^{-1}{\bf F}\,\partial_\bp )^mf_F~.
\label{g}
\ee
A useful way of visualizing this solution of Eq.~(\ref{1}) at given
order in ${\cal E}_{\rm dc}$ and ${\cal E}_\omega$ is by counting all
possible couplings of harmonics $g_{\nu n}$ represented as points on
the $(\nu, n)$ plane (Fig.~\ref{f1}). The {\it dc} field ${\cal E}_{\rm dc}$ 
couples
nearest-neighbor harmonics along the $\nu$ axis, $g_{\nu
n}\leftrightarrow g_{\nu\pm 1,n}$. We are interested here in the
linear (with respect to ${\cal E}_{\rm dc}$) photoconductivity
$\sigma_{\rm ph}$, so that only one such link is allowed. The {\it ac} field
${\cal E}_\omega$ couples harmonics along diagonals, $g_{\nu
n}\leftrightarrow g_{\nu\pm 1,n\pm 1}$ and $g_{\nu\pm 1,n\mp 1}$. 
The propagator $L^{-1}$ is a diagonal matrix in $(\nu, n)$ space.
The static longitudinal current $j=\sigma_{\rm ph}{\cal E}_{\rm dc}$,
\be
j=-{e\over 2\pi\hbar^2}\int\!d\ve\,p\,{\rm Re}\,g_{10}~,
\label{j}
\ee 
is expressed through $g_{10}$, i.e., is given by a sum of all paths
starting at (0,0) and ending at $(1,0)$. Already at order $\sigma_{\rm
ph}\sim{\cal O}({\cal E}_{\rm dc}^0{\cal E}_\omega^2)$ as many as 9
different graphs arise, shown in Fig.~\ref{f1}, to which one should
add their counterparts mirrored in the horizontal axis, which
corresponds to the change $\omega\to -\omega$.

Our strategy for finding $g_{10}$ is to select graphs involving
couplings whose strength diverges at $\tau_{\rm in}/\tau\to\infty$.
These are graphs returning to the point $(0,0)$ [graphs (a), (b) in
Fig.~\ref{f1}], which are proportional to $(L^{-1})_{00}$. This means
that to order ${\cal E}_\omega^2$ the path $(0,0)\to (1,1)\to (0,0)\to
(1,0)$ (and its counterparts in other quadrants) gives the main
contribution to $\sigma_{\rm ph}$ for 
\begin{equation}
\tau_{\rm in}\gg\tau~.
\end{equation}
The perturbative expansion in powers of ${\cal E}_\omega^2$ proceeds by
iterating the loop $(0,0)\to (1,1)\to (0,0)$.  In this way we arrive at a
simple relation $(\pm i\omega_c+\tau^{-1})g_{\pm 1,0}=e{\cal E}_{\rm
dc}\,\partial_pf/2$, where $f=g_{00}$ satisfies the closed equation
\be 
{e^2{\cal E}_\omega^2\over
2m}\,\partial_\varepsilon\!\left[ K_\omega \,\partial_\varepsilon
f\right] + I_{\rm in}f=0~. \label{3} 
\ee 
The function $K_\omega=K_\omega^{+}+K_\omega^{-}$ describes the
absorption rate at energy $\varepsilon$: 
\be
2K^{\pm}_\omega={\varepsilon\tau\over
1+(\omega\pm\omega_c)^2\tau^2}~. \label{4} 
\ee 
The photoconductivity at $\tau_{\rm in}\gg\tau$ is thus completely
determined by $f$, i.e., in this limit the {\it ac} field modifies
the {\it dc} current through the heating: 
\be 
\sigma_{\rm ph}=-{e^2\over
2\pi\hbar^2}\int \!\!d\varepsilon\, K_0\,\partial_\varepsilon
f~.\label{5} 
\ee

The function $f(\varepsilon)$ changes abruptly around the Fermi energy
$\epsilon_F$ on a scale 
\be
T_e=\max\{T,\Delta_h\},
\ee
 where
\be
\Delta_h=(\,e{\cal E}_\omega\,l_{\rm
in}/2\,)\,[\,K_\omega(\epsilon_F)/K_0(\epsilon_F)
\,]^{1/2}
\label{6}
\ee 
and $l_{\rm in}=v_F\,[\,K_0(\epsilon_F)\tau_{\rm
in}/\epsilon_F\,]^{1/2}$ is the inelastic length ($v_F$ is the
Fermi velocity). Note
that $\tau_{\rm in}$ in the regime of strong heating ($\Delta_h\gg T$)
should be found self-consistently
with $\Delta_h$ and thus depends on ${\cal E}_\omega$.

Turning to the evaluation of $\sigma_{\rm ph}$ under the assumption
that $T_e\ll \epsilon_F$, we first notice that a seemingly reasonable
approximation which neglects the $\varepsilon$ dependence of $\tau$ in
the integrand of Eq.~(\ref{5}) [normally, $\tau(\varepsilon)$ changes
on a scale of $\epsilon_F$] yields an identically zero
photoresponse. Indeed, in that case $\sigma_{\rm ph}$ is equal to the
static Drude conductivity $\sigma^{\rm D}_0$ independently of the
detailed shape of $f(\varepsilon)$, since $\int d\varepsilon
\,\varepsilon\,\partial_\varepsilon f=-2\pi\hbar^2n_e/m$ due to
particle number conservation ($n_e$ is the electron concentration). It
follows that the dependence of $\tau$ on $\varepsilon$ should be taken
into account. It is worth mentioning that, contrary to a naive
expectation, this does not lead to any additional 
smallness
of $\sigma_{\rm ph}$ 
since $\tau(\ve)$ enters the result through expressions of the type
$\ve\partial_\ve\tau|_{\ve=\ve_F}\sim\tau$.

We do not discuss specific microscopic models of the inelastic
coupling of electrons to a thermal bath, our purpose here is to use
the simplest possible representation of $I_{\rm in}$. In a conserving
relaxation-time approximation
\be I_{\rm in}f=-\tau_{\rm in}^{-1}(f-f_F)~, 
\label{7} 
\ee 
where the $\varepsilon$ independent $\tau_{\rm in}^{-1}$ is in general
a functional of $f(\varepsilon)$, we get from Eqs.~(\ref{3}),(\ref{5}) for
$T_e\ll\epsilon_F$:
\begin{eqnarray}
\sigma_{\rm ph}-\sigma^{\rm D}_0&=&\sigma^{\rm
D}_\omega\,{e^2{\cal E}_\omega^2\tau_{\rm in}\over 2m}\,\,K''_0
\nonumber \\
&=&{e^2\over 2\pi\hbar^2}\,\,\Delta_h^2K''_0~.
\label{8}
\end{eqnarray} 
Here $K''_0=\partial^2_\varepsilon K_0|_{\varepsilon=\epsilon_F}$ and
$\sigma_\omega^{\rm D}$ is the zero-$T$ dynamic Drude conductivity,
$\sigma^{\rm D}_\omega=e^2K_\omega (\epsilon_F)/2\pi\hbar^2$. Note
that the only source of nonlinearity of $\sigma_{\rm ph}$ with respect
to the {\it ac} field power in Eq.~(\ref{8}) is a dependence of
$\tau_{\rm in}$ on ${\cal E}_\omega$.

Alternatively, assuming the dominant role of soft inelastic scattering
with energy transfers much smaller than $T$, we can write $I_{\rm in}$
in the Fokker-Planck form:
\be
I_{\rm
in}f=\partial_\varepsilon\,\{B\,[\,\partial_\varepsilon
f+T^{-1}f(1-f)\,]\}~, \label{9}
\ee 
where
$B(\varepsilon)=\left<(\delta\varepsilon)^2
W(\varepsilon,\delta\varepsilon)\right>/2$
is the diffusion coefficient in energy space, $W$ is the corresponding rate of
inelastic processes, and $\left<\ldots\right>$ denotes averaging over the
energy transfer $\delta\varepsilon$. Equation (\ref{3}) becomes then
first-order in $\partial_\varepsilon$, which gives $f(\varepsilon)$ described
by the Fermi distribution with the effective electron temperature
\be
T_{\rm eff}=T+\Delta_{\rm
FP}, 
\ee
where 
\be
\Delta_{\rm FP}=e^2{\cal
E}_\omega^2TK_\omega(\epsilon_F)/2mB(\epsilon_F)~,
\label{9a}
\ee and
\begin{eqnarray}
\sigma_{\rm ph}-\sigma_0^{\rm D}&=&{e^2\over
2\pi\hbar^2}\,{\pi^2\over 6}\,(T_{\rm eff}^2-T^2)\,K''_0~
\nonumber \\&=&{e^2\over
2\pi\hbar^2}\,{\pi^2\over 6}\,(2T+\Delta_{\rm FP})\Delta_{\rm FP}K''_0~.
\label{10} \end{eqnarray}
The case of typical energy transfers $\sim T$ may be
qualitatively described by either model with
$B(\epsilon_F)\tau_{\rm in}\sim T_eT$.

The microwave power and temperature dependences of $\sigma_{\rm ph}$
can be found from Eqs.~(\ref{8}),(\ref{10}) for a variety of
scattering mechanisms. If one assumes that $\tau_{\rm in}$ is
determined by scattering on acoustic phonons via the piezoelectric
interaction screened by the 2DEG, the characteristic energy transfer
is $T_e$ and $\tau_{\rm in}^{-1}\propto T_e^3$. It follows then from
Eq.~(\ref{6}) that the heating at $\Delta_h\gg T$ is characterized by
$T_e\propto {\cal E}_\omega^{2/5}$. By using Eq.~(\ref{8}) we get
$\sigma_{\rm ph}-\sigma^{\rm D}_0\propto {\cal E}_\omega^2T^{-3}$ for
$\Delta_h\ll T$ and $T$ independent $\sigma_{\rm ph}-\sigma^{\rm
D}_0\propto {\cal E}_\omega^{4/5}$ otherwise.

Having identified the main contribution to $\sigma_{\rm ph}$ in the 
limit $\tau_{\rm in}/\tau\gg1$  [diagrams (a),(b) in Fig.~\ref{f1}]
it is instructive to compare this contribution with that corresponding to
 other diagrams
[diagrams (c)--(i)]. While the former is related to the 
heating of electrons by the {\it ac} field, the latter can be regarded
as an effect of radiation on the impurity scattering and 
thus represents a classical analog of the quantum effect considered in 
Refs.~\onlinecite{durst03}--\onlinecite{vavilov03}.
Following the procedure given by Eqs.~(\ref{g}), (\ref{j}) and 
making use of the explicit matrix form of the field operator,
$[{\bf F}\partial_\bp
g]_{\nu n}=F^{\nu\nu'}_{n n'}g_{\nu' n'}$,
\bea\nonumber
&&\!\!F^{\nu\nu'}_{n n'}=-{1\over2}\delta_{\nu,\nu'\pm1}
\left({\bf s}_{\nu\nu'}\cdot e\vec{\cal E}_{nn'}\right)
\left[\partial_p+(\nu'-\nu){\nu'\over p}\right]~,\\\nonumber
&&\!\!\vec{\cal E}_{nn'}={\cal \vec{E}}_{\rm dc}\delta_{nn'}+
{1\over2}{\cal \vec{E}}_\omega\delta_{n,n'\pm 1},\\
&&{\bf s}_{\nu\nu'}={\bf e}_x+i(\nu'\!-\nu){\bf e}_y~,
\label{F}
\eea
one can readily calculate the photoconductivity at any desirable 
order in the fields ${\cal \vec{E}}_{\rm dc}$ and ${\cal \vec{E}}_\omega$
(${\bf e}_{x,y}$ are the unit vectors along the $x,y$ axes).
At the lowest order the photoconductivity $\sigma_{\rm
ph}\sim{\cal O}({\cal E}_{\rm dc}^0{\cal E}_\omega^2)$ is given 
 by the diagrams (a)--(i) in Fig.~\ref{f1} 
(together with their counterparts in a
lower half-plane, $\omega\to-\omega$). The result takes a simple form
in the limit $\tau^{-1}_{\rm in}\ll\tau^{-1}\ll\omega_c\ll\omega$:
\bea
&&\sigma_{\rm ph}-\sigma^{\rm D}_0={1\over8}\sigma^{\rm D}_0
\left(\frac{eE_\omega v_F}{\ve_F\omega}\right)^2\\\nonumber
&&\times\left[\,2c_1{\tau_{\rm in}\over\tau}+(5c_1+4c_2)
+(3c_1+2c_2)
\cos2\phi_E\,\right],
\label{all}
\eea
where $c_1=\ve\tau\partial^2_\ve\ve\tau^{-1}|_{\ve=\ve_F}$, 
$c_2=\ve\tau\partial_\ve\tau^{-1}|_{\ve=\ve_F}$ are numbers (typically of 
order unity) determined by the type of disorder, $\phi_E$ is the angle 
between ${\cal \vec{E}}_{\rm dc}$ and ${\cal \vec{E}}_\omega$. 
The first term in the square
brackets corresponds to the diagrams (a), (b) in Fig.~\ref{f1} and reproduces 
Eq.~(\ref{8}) in the limit of weak heating, $\Delta_h\ll T$.
The term $(5c_1+4c_2)$ corresponds to the diagrams (c), (e), and (f).
The polarization--dependent part, given by the last term, 
originates from the diagrams (d) and (i) (in which both diagonal 
links have
the same direction along the $\nu$ axis). Finally, the diagrams (g) and (h)
give a contribution which is smaller, compared to the diagrams
(c)--(f) and (i), in the parameter $1/\omega_c\tau_{\rm in}\ll 1$
and is omitted in Eq.~(\ref{all}). One can clearly see 
from Eq.~(\ref{all}) that in the limit $\tau_{\rm in}/\tau\gg 1$ 
the photoconductivity is dominated by the heating of electrons.

\section{Interaction-induced photoconductivity}
\label{sec3}

In the above, we have neglected inelastic electron-electron
collisions, whose role is not essential for the quasiclassical OPC,
but have also ignored the renormalization of the {\it elastic}
scattering rate by electron-electron interactions. The latter
approximation, which fits in with the conventional approach to the
photoconductivity, in fact misses an important contribution to
$\sigma_{\rm ph}$. Recall that the change of the conductivity due to
radiation at $\tau_{\rm in}\gg\tau$ comes mainly from the heating. It
is most illuminating to focus on the model of
Eqs.~(\ref{9}),(\ref{10}), within which $\sigma_{\rm ph}-\sigma^{\rm
D}_0$ is simply proportional to $T_{\rm eff}^2-T^2$. Clearly, this
contribution to $\sigma_{\rm ph}$ is associated with the term in the
Drude conductivity that is quadratic in the small parameter
$T/\epsilon_F$. Substituting $T_{\rm eff}$ for $T$ in the Drude term
yields $\sigma_{\rm ph}$ given by Eq.~(\ref{10}). On the other hand,
there are $T$ dependent quantum corrections to the conductivity,
neglected above, in which one should similarly change $T\to T_{\rm
eff}$. At low $T$, the terms in $\sigma_{\rm ph}$ coming from these
quantum corrections may easily become larger than the classical
contribution (\ref{10}), as we now demonstrate.

For high-mobility samples, we are mostly interested in $\sigma_{\rm
ph}$ at not too low temperatures $T\tau/\hbar\gg 1$. In this
``ballistic" regime, the most important $T$ dependent term in the
conductivity at zero $B$, for the limiting case of short-range
disorder ($\tau\to\tau_S$), is related to screening of the disorder by
Friedel oscillations, which translates into a $T$ and $\varepsilon$
dependent renormalization of the elastic scattering rate. This
quantum interaction-induced term is given by $\Delta\sigma_{\rm
int}=\alpha (e^2/\pi\hbar^2)T\tau_S$. \cite{zala01} Here $\alpha$ is
the interaction coupling constant, equal to unity for the Coulomb
interaction (under the assumption that $k_F^{-1}$ is much smaller than
the static screening length). Remarkably, $\Delta\sigma_{\rm int}$ is
linear in $T/\epsilon_F$, in contrast to the classical $T$ dependent
term which is quadratic in $T/\epsilon_F$. Substitution $T\to T_{\rm
eff}$ in $\Delta\sigma_{\rm int}$ yields an interaction-induced term
in the photoconductivity, $\Delta\sigma_{\rm ph}=\alpha
(e^2/\pi\hbar^2)(T_{\rm eff}-T)\tau_S$, where $T_{\rm
eff}-T=\Delta_{\rm FP}$ is given by Eq.~(\ref{9a}). For finite $B$,
assuming that $T\gg \hbar\omega_c, \hbar/\tau$, this term in $\sigma_{\rm
ph}$ reads
\be 
\Delta\sigma_{\rm ph}=\alpha \,{e^2\over \pi\hbar^2}\,(T_{\rm
eff}-T)\tau_S\,{1-\omega_c^2\tau_S^2\over (1+\omega_c^2\tau_S^2)^2}~. 
\label{10a}
\ee
This result is obtained by inverting the resistivity tensor for which the
leading (for $T\gg \hbar\omega_c, \hbar/\tau$) interaction-induced correction
to $\rho_{xx}$ is $B$ independent, while that to $\rho_{xy}$ may be
neglected. It follows from the comparison of Eqs.~(\ref{10}),(\ref{10a}) that
this quantum contribution to the photoconductivity is much larger than the
classical one provided the effective temperature is low, $T_{\rm eff}\ll
\alpha\epsilon_F$, which is satisfied for $\alpha\sim 1$ in the whole range of
temperatures in a degenerate Fermi system. Thus sufficiently strong
interactions have the effect of greatly enhancing the photoconductivity.

For stronger magnetic fields, $\hbar\omega_c\gg T$, another mechanism of the
interaction-induced photoconductivity becomes relevant, related to the
interplay \cite{gornyi03} of quasiclassical memory effects and
electron-electron interactions. For the two-component model of disorder,
assuming, as above, that $\tau_S\ll\tau_L$, the $T$ dependent correction to
the conductivity is $\Delta\sigma_{\rm int}\sim \alpha
(e^2/\hbar)(\tau_L/\tau_S)^{1/2}(T\tau_S/\hbar)^{-1/2}$. \cite{gornyi03} With
numerical factors included, this yields a contribution to the
photoconductivity
\be
\Delta\sigma_{\rm ph}=-\alpha{e^2\over
\pi\hbar}{3\zeta(3/2)\over 16\pi^{3/2}}\left({\tau_L\over
\tau_S}\right)^{1/2}{T_{\rm eff}^{-1/2}-T^{-1/2}\over (\tau_S/\hbar)^{1/2}}~.
\label{10b}
\ee

Comparing Eqs.~(\ref{10a}) and (\ref{10b}), one sees that the latter
mechanism gives a larger contribution to $\sigma_{\rm ph}$ in the
whole temperature range $T\alt\hbar\omega_c$. At $T\sim\hbar\omega_c$,
the term (\ref{10b}) is still larger than that given by
Eq.~(\ref{10a}) by a factor $(\omega_c\tau_L)^{1/2}\gg 1$. With
increasing $T$, however, the memory-effects induced correction falls
off rapidly, as $\exp [-4\pi^2T/\omega_c]$, so that at
$T\sim\hbar\omega_c\ln(\omega_c\tau_L)$ a crossover to
Eq.~(\ref{10a}) occurs.

\section{Magnetooscillations due to memory effects}
\label{sec4}

The photoconductivity obtained in Secs.~\ref{sec2},\ref{sec3} 
exhibits the cyclotron resonance
but shows no oscillations with varying $\omega/\omega_c$. Let us now
incorporate the memory effects. To this end, we have to step back 
to write the Liouville equation not yet averaged over
positions of impurities. More precisely, for the two-component model of
disorder (specified in Sec.~\ref{sec1}), we average only over
smooth disorder and represent the Liouville operator as
\be 
L=L_0+\delta
L-I_{\rm in}~, 
\label{11}
\ee
where $L_0$ includes the effect of scattering on
smooth disorder:
\be
L_0=\partial_t+{\bf v}\nabla_{\bf
r}+\omega_c\,\partial_\phi-\tau_L^{-1}\partial^2_\phi~,
\label{12}
\ee
and $\delta L=-\sum_iI_{{\bf R}_i}({\bf r})$ is a sum of collision operators
for short-range impurities located at points ${\bf R}_i$. We have to keep in
$L_0$ the spatial-gradient term (${\bf v}$ is the velocity).

Averaging the solution of Eq.~(\ref{1}) over ${\bf R}_i$ with $L$ given by
Eq.~(\ref{11}) can be done systematically along the lines of
Ref.~\onlinecite{mirlin01}: a classical diagram technique is formulated by
means of the free propagator $L_0^{-1}$ and the disorder correlation function
$\left<\delta L({\bf r})\delta L({\bf r}')\right>$.  We proceed by
representing the averaged propagator $\left<L^{-1}\right>=(L_0+M-I_{\rm
in})^{-1}$ in terms of the self-energy operator $M$. Equations
(\ref{3})--(\ref{5}) are then reproduced with $\tau$ in Eq.~(\ref{4}) given by
\be
\tau^{-1}=\tau_L^{-1}+\Sigma (\omega)~,
\label{12a}
\ee 
where $\Sigma=\int
(d\phi/2\pi) \,{\bf n}M{\bf n}$ and ${\bf n}={\bf v}/|{\bf v}|$. To
first order in $\delta L$, 
\be
\Sigma^{(1)}=-n_S\int \!d{\bf r}\!\int
(d\phi/2\pi) \,{\bf n} I_{{\bf R}_i}({\bf r}){\bf n}~. 
\label{12b}
\ee 
By definition $n_S\!\int \!d{\bf r}\, I_{{\bf R}_i}({\bf r}){\bf
n}=-{\bf n}\tau_S^{-1}$, so that we have $\Sigma^{(1)}=\tau_S^{-1}$,
which yields the Drude result for the total scattering rate
$\tau^{-1}=\tau_L^{-1}+\tau_S^{-1}$. Expanding now $M$ to second order
in $\delta L$, we obtain the leading correction to $\Sigma$ that is
due to the memory effects:
\begin{eqnarray}
\Sigma^{(2)}(\omega)&=&-n_S\!\int \!\!d{\bf r}\!\int\!\!
d{\bf r}'\!\!\int\!
{d\phi\over 2\pi} \nonumber \\
&\times&{\bf n}\,I_{{\bf R}_i}({\bf r})\,D_\omega({\bf r}-{\bf
r}')\,I_{{\bf R}_i}({\bf r}')\,{\bf n}~, \label{13}
\end{eqnarray}
or more explicitly
\begin{eqnarray}
\Sigma^{(2)}(\omega)&=&-4\pi n_S\!\int \!\!d{\bf r}\!\int\!\!
d{\bf r}'\!\!\int\!
{d\phi\over 2\pi} \!\!\int\!
{d\widetilde{\phi}\over 2\pi}\!\!\int\!
{d\widetilde{\phi}'\over 2\pi}\!\!\int\!
{d\phi'\over 2\pi}\nonumber \\
&\times&\cos\phi\,I_{{\bf R}_i}({\bf r},\phi,\widetilde{\phi})\,D_\omega({\bf r}-{\bf
r}',\widetilde{\phi},\widetilde{\phi}')\nonumber\\
&\times&I_{{\bf R}_i}({\bf r}',\widetilde{\phi}',\phi')\,\cos\phi'~,
\label{13a}
\end{eqnarray}
where the propagator 
\be
D=(L_0+\tau_S^{-1})^{-1}
\label{D}
\ee
is taken in the $\omega$ representation. Most importantly, the $\omega$
dispersion of $D$ leads to oscillations of $\Sigma^{(2)}(\omega)$ with a
period $\omega_c$.

To find $\Sigma^{(2)}(\omega)$, we first note that, since $I_{{\bf R}_i}({\bf
r})$ as a function of ${\bf r}$ falls off fast beyond a small vicinity of
${\bf R}_i$, one can put ${\bf r}={\bf r}'$ in the argument of $D_\omega$ in
Eqs.~(\ref{13}),(\ref{13a}). Then, $\Sigma^{(2)}(\omega)$ is given by
\begin{eqnarray}
\Sigma^{(2)}(\omega)&=&-{2\over n_S\tau_S^2}\int\!{d\phi\over
2\pi}\!\int\!d\phi' \nonumber\\ &\times&\cos\phi
\,D_\omega(0;\phi,\phi')\,\cos\phi'~,
\end{eqnarray}
where $D_\omega(0;\phi,\phi')$ is the Fourier transform in $t$ of the
probability density to return with a direction of ${\bf v}$ specified by
$\phi'$ if one starts at an angle $\phi$.  Let us now focus on the case
$\omega_c\tau_S\gg 1$. In this limit, $D_\omega(0;\phi,\phi')$ is sharply
peaked at $\phi=\phi'$ and, introducing the total probability of return
$P_\omega=\int\!d\phi'D_\omega(0;\phi,\phi')$, we finally get
\be
\Sigma^{(2)}(\omega)=-P_\omega/n_S\tau_S^2~.\label{14}
\ee 

A return-induced correction to the effective scattering rate, which
comes according to Eq.~(\ref{14}) from ${\rm Re}\,P_\omega$,
yields, away from the cyclotron resonance, oscillations of
the absorption rate through a correction to the function
$K_\omega(\epsilon_F)$ [cf.\ Eq.~(\ref{4})]:
\be
\Delta K^\pm_\omega(\epsilon_F)=-(\epsilon_F/2n_S\tau_S^2)\,{\rm
Re}\,P_\omega/(\omega\pm\omega_c)^2~.
\label{14a}
\ee
The oscillatory part of $K_\omega(\epsilon_F)$ leads to {\it
classical} oscillations of the linear {\it ac} conductivity,
\cite{polyakov02} 
\be
\Delta\sigma_\omega^{(c)}=-\sigma_\omega^{\rm
D}\,{\rm Re}\,P_\omega/n_S\tau_S, 
\label{20-1}
\ee
and being substituted in
Eqs.~(\ref{3})--(\ref{5}), to those of $\sigma_{\rm ph}$.  To first
order in ${\cal E}_\omega^2$, the classical oscillatory correction to
$\sigma_{\rm ph}$ reads
\be
{\Delta\sigma_{\rm ph}^{(c)}\over \sigma_{\rm ph}-\sigma_0^{\rm
D}}={\Delta\sigma_\omega^{(c)}\over \sigma_\omega^{\rm D}}=-{{\rm
Re}\,P_\omega\over n_S\tau_S}~.\label{20}
\ee
It is worth noting once more that both the smooth correction
$\sigma_{\rm ph}-\sigma_0^{\rm D}$ and the oscillatory contribution
$\Delta\sigma_{\rm ph}^{(c)}$ are proportional to the inelastic time
$\tau_{\rm in}$.

\begin{figure}
\centerline{ 
\includegraphics[width=8cm]{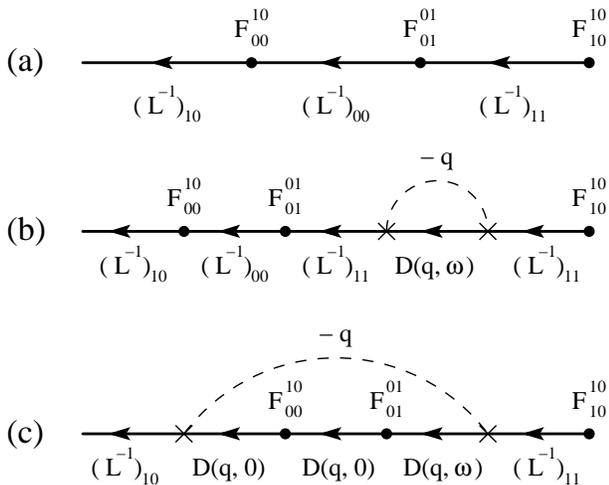}}
\vspace{3mm}
\caption{
   Diagrams describing the memory effects in the photoconductivity
  $\sigma_{\rm ph}$: oscillatory self-energy (b) and vertex (c)
  corrections to the smooth part (a) of $\sigma_{\rm ph}$. }
\label{f2}
\end{figure}

In the above, we have analyzed the oscillatory correction to the
self-energy in terms of the return probability $P_\omega$. In fact,
there are other contributions to the OPC which are not reduced to the
self-energy corrections and cannot be represented through $P_\omega$.
To illustrate this point, it is convenient to switch to a more
conventional (dual) representation of the diagrams in Fig.~\ref{f1},
now with lines corresponding to the propagators and vertices
representing the field operators (\ref{F}), as shown in Fig.~\ref{f2}.
The diagram (a) in Fig.~\ref{f2} reproduces the graph (a)
in Fig.~\ref{f1}. The diagram Fig.~\ref{f2}(b) represents the
oscillatory correction to $\sigma_{\rm ph}$ of the self-energy type,
Eq.~(\ref{20}). Both diagrams (a) and (b) in Fig.~\ref{f2} contain the
inelastic propagator $(L^{-1})_{00}=\tau_{\rm in}$ at zero momentum
$q$, which is much larger than all other propagators,
$(L^{-1})_{\nu n}$ with at least one of the indices $\nu, n\neq 0$.  
By contrast, the diagram (c),
which exemplifies an oscillatory vertex correction to $\sigma_{\rm
  ph}$, is not proportional to $\tau_{\rm in}$, because of large $q$
running along the internal propagators $D(q,\omega)$ [defined in
Eq.~(\ref{D})].  The vertex type corrections, which are of the same
order in all of the diagrams (a)--(i), are thus by a factor $\tau_{\rm
  in}/\tau_S$ smaller than the self-energy contribution (\ref{20}).

The function $P_\omega$ for $\omega_c\tau_L\gg 1$ is most directly
evaluated by using Eq.~(\ref{12}) which represents the time evolution
of $\phi$ as a diffusion process with a white noise spectrum of
$\partial_t\phi$. This approach is justified for not too strong $B$,
namely for $\delta\gg d$, where (see Appendix)
\be
\delta=2\pi^{1/2}v_F\tau_L/(\omega_c\tau_L)^{3/2}
\label{delta}
\ee
is a mean-square fluctuation of the guiding center of a cyclotron orbit after
one cyclotron revolution (otherwise adiabatic drift dynamics is
developed). The probability density $p_n(x_\perp,x_\parallel)$ for particles
on the Fermi surface to be scattered from the starting point on a cyclotron
orbit by a distance $x_\perp$ across the orbit and a distance $x_\parallel$
along it in time $2\pi n/\omega_c$ is then given by the anisotropic Gaussian
distribution with averages
$\left<x^2_\perp\right>=\left<x^2_\parallel\right>/3=n\delta^2/2$ (see
Appendix). 
 Summing over multiple cyclotron revolutions, we thus express
$P_\omega$ as
\be
P_\omega=\sum\limits_{n=1} \int\limits_{-\infty}^\infty \! dt \,
e^{-i(\omega-i/\tau_S)t}\,p_n[\,0,v_F(t-2\pi n/\omega_c)\,]~.\label{15} 
\ee
Note that once the particle hits
a short-range impurity, its guiding center is shifted by a distance of the order of the
cyclotron radius. As a result, the contribution of such
trajectories to the return probability can be neglected and only non-colliding
orbits should be taken into account, 
which is expressed by the exponential factor $\exp(-t/\tau_S)$. 
Equation (\ref{15}) gives oscillations of $P_\omega$ as $\omega/\omega_c$ is varied:
\be
P_\omega={1\over
\sqrt{\pi}v_F\delta}\sum\limits_{n=1}{1\over
\sqrt{n}}\exp
\left[-{2\pi n\over\omega_c}\left(i\omega+\Gamma\right)\right]~,
\label{16}
\ee
whose damping with decreasing $B$ is characterized by
\be 
\Gamma={3\over
2\tau_L}\left({\omega\over\omega_c}\right)^2 + {1\over \tau_S}~.
\label{17}
\ee

In the limit of weak damping, $\pi\Gamma\ll\omega_c$, we perform the
summation in Eq.~(\ref{16}) by means of Poisson's formula to represent
${\rm Re}\,P_\omega$ as a series of sharp peaks centered at
$\omega=N\omega_c$. A peak at $\omega\simeq N\omega_c$ is of the form
\cite{rudin97}
\begin{eqnarray} 
{\rm Re}\,P_\omega&=&{\omega_c^3\tau_L\over 2\sqrt{3}\pi
v_F^2\omega}\,{\cal F}\left({\omega-N\omega_c\over
\Gamma}\right)~,\label{18a}\\ {\cal
F}(x)&=&\left[{1+(1+x^2)^{1/2}\over 2(1+x^2)}\right]^{1/2}~,\quad
{\cal F}(0)=1~.
\label{18}
\end{eqnarray}
Note that the amplitude of the peaks in Eq.~(\ref{18a}) falls off with
decreasing $\omega_c$ or increasing $\omega$ as a power law, namely as
$\omega_c^3/\omega$. The power-law suppression of the oscillations
crosses over into the exponential damping only for very large $\pi
(\omega/\omega_c)^2\gg\omega_c\tau_L$, when one can neglect all terms
in Eq.~(\ref{16}) but the first one, which gives
\be
{\rm Re}\,P_\omega={(\omega_c\tau_L)^{3/2}\over 2\pi
v_F^2\tau_L}\cos{2\pi\omega\over \omega_c}\,\exp
\left(-{2\pi\Gamma\over\omega_c}\right)~. \label{19}
\ee
It is worth noting that, because of the condition $\omega_c\tau_S\gg 1$, the
term $\tau_S^{-1}$ in Eq.~(\ref{17}) may be neglected in the damping factor of
Eq.~(\ref{19}), so that the exponential damping is determined by the momentum
relaxation time for scattering off the long-range disorder.

\section{Oscillatory ac conductivity: Quasiclassical vs quantum}
\label{sec5}

Now we compare the classical oscillatory ac conductivity $\sigma_\omega^{(c)}$,
given by Eqs.~(\ref{20-1}),(\ref{18a}),(\ref{19}), with the quantum
contribution $\sigma_\omega^{(q)}$ calculated in
Ref.~\onlinecite{dmitriev03}. Let us represent
$\sigma_\omega^{(c)}$ for weak damping at $\omega=N\omega_c$ as
\be
\left.\sigma_\omega^{(c)}\right|_{\,\omega=N\omega_c}=\sigma_\omega^{\rm
    D}\left[1-{a\over\sqrt{3\pi}\,N\delta}\,(\omega_c\tau_L)^{1/2}\right]~,
\ee 
where $\delta$ is given by Eq.~(\ref{delta}) and
we have introduced a characteristic size of the short-range impurities
$a=(n_Sv_F\tau_S)^{-1}$. It follows that, apart from the harmonics number $N$, 
the amplitude of the oscillations is given by the product of 
a small factor $a/\delta$ and a large factor $(\omega_c\tau_L)^{1/2}$.
In the exponential damping regime, $\sigma_\omega^{(c)}$ is re-written 
as
\bea
{\sigma_\omega^{(c)}\over \sigma_\omega^{\rm D}}=
1-{a\over\sqrt{\pi}\,\delta}\cos\frac{2\pi \omega}{\omega_c}
\exp\left[-\left({\omega\over\omega_c}\right)^2{3\pi\over
\omega_c\tau_L}\right]~,
\label{ac_cl}
\eea
so that the pre-exponential factor is simply given by $a/\delta$. An
important point to notice is that the damping in Eq.~(\ref{ac_cl}) is
characterized solely by the long transport time for scattering off the
smooth disorder.  On the other hand, the envelope of the quantum
oscillations of the {\it ac} conductivity is determined by the
single-particle time $\tau_q$:
\be
{\sigma_\omega^{(q)}\over \sigma_\omega^{\rm D}}=
1+2\cos\frac{2\pi \omega}{\omega_c}
\exp\left(-\frac{2\pi}{\omega_c \tau_{\rm q}}\right)
\label{ac_q}
\ee
(this equation is valid for $2\pi T\gg \hbar/\tau_q$, for smaller $T$
see Ref.~\onlinecite{dmitriev03}). 

\begin{figure}
\centerline{ 
\includegraphics[width=8cm]{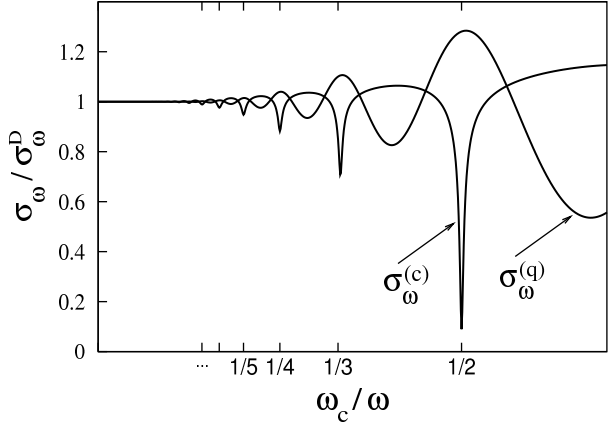}}
\vspace{3mm}
\caption{
Quasiclassical [$\sigma_\omega^{(c)}$, Eq.~(\ref{ac_cl})] 
and quantum [$\sigma_\omega^{(q)}$, Eq.~(\ref{ac_q})]
oscillatory  {\it ac} conductivity (normalized to
the Drude conductivity  $\sigma_\omega^{\rm D}$) vs $\omega_c/\omega$ for
$\omega/2\pi=100$~GHz, $\tau=0.6$~ns, $\tau/\tau_{\rm q}=50$, 
$\tau_S/\tau_L=0.1$,
$a/\delta=0.25$ at $\omega_c/\omega=1/2$. 
}
\label{f3}
\end{figure}

Note the difference in the sign of
the oscillatory terms: there is a $\pi$ shift of the quantum and
classical oscillations with respect to each other. Another difference
is that the damping of the classical oscillations is $\omega$
dependent, in contrast to the quantum case.  One sees that, despite
the small factor $a/\delta$, the classical oscillations may be 
stronger than the quantum ones since
in high-mobility structures $\tau_q\ll\tau_L$ and the quantum
oscillations are damped much more strongly. The behavior of the two
contributions to the oscillatory {\it ac} conductivity is illustrated
in Fig.~\ref{f3}.

\section{Mechanisms of the oscillatory photoconductivity: 
Quasiclassical vs quantum}
\label{sec6}

Having found the classical contribution $\Delta\sigma_{\rm ph}^{(c)}$
to the OPC [Eqs.~(\ref{20}),(\ref{18a}),(\ref{19})], let us compare it
with the quantum oscillatory contribution $\Delta\sigma_{\rm
ph}^{(q)}$, \cite{dmitriev03,dmitriev03a} related to the oscillations
of the DOS. Using Eqs.~(\ref{20}),(\ref{17}),(\ref{19}) and omitting
numerical factors, we write down the essential factors in
$\Delta\sigma_{\rm ph}^{(c)}$ for $\pi\Gamma\agt \omega_c$ for the
case of noninteracting electrons:
\begin{eqnarray}
\Delta\sigma_{\rm ph}^{(c)}&\sim& \sigma_0^{\rm D}\,{\tau_{\rm
in}\over\tau}\,\left({e{\cal E}_\omega v_F\over
\epsilon_F\omega}\right)^2 {a\over\delta}\nonumber\\ &\times&\cos
{2\pi\omega\over\omega_c}\,\exp \left[
-\left({\omega\over\omega_c}\right)^2{3\pi\over
\omega_c\tau_L}\right]~.
\label{21a}
\end{eqnarray}
The sign of $\Delta\sigma_{\rm
ph}^{(c)}$ in Eq.~(\ref{21a}) depends on that of $K_0''$ [see
Eq.~(\ref{8})].

As shown in Sec.~\ref{sec3}, unless the electron-electron interaction
is extremely weak, the largest contribution to the smooth part of
$\sigma_{\rm ph}$ comes from the interaction correction to the
conductivity. For $T\agt \hbar\omega_c\ln(\omega_c\tau_L)$, the
main interaction-induced term in the smooth part of $\sigma_{\rm ph}$
is given by Eq.~(\ref{10a}) and, according to Eq.~(\ref{20}), this yields
in turn the main term in the oscillating part $\Delta\sigma^{(c)}_{\rm
ph}$. Putting $\alpha\sim 1$ (long-range Coulomb interaction), we
have: 
\begin{eqnarray} 
\Delta\sigma_{\rm ph}^{(c)}&\sim& \sigma_0^{\rm
D}\,{\tau_{\rm in}\over\tau}\,{\left({e{\cal
E}_\omega v_F/\omega}\right)^2\over \epsilon_F T}\,
{a\over\delta}\nonumber\\ &\times&\cos {2\pi\omega\over\omega_c}\,\exp
\left[ -\left({\omega\over\omega_c}\right)^2{3\pi\over
\omega_c\tau_L}\right]~, 
\label{21} 
\end{eqnarray} 
which is larger than the noninteracting
part [Eq.~(\ref{21a})] by a factor $\epsilon_F/T$.

For $T\ll\hbar\omega_c$, the main contribution to $\sigma_{\rm ph}$ is
related to the interaction correction given by Eq.~(\ref{10b}), which
yields the oscillatory part $\Delta\sigma_{\rm ph}^{(c)}$ similar (in
terms of the phase of the oscillations and their damping factor) to
that in Eq.~(\ref{21}) but multiplied by a large factor
$(\omega_c\tau_L)^2/(T\tau_L/\hbar)^{3/2}$. In the intermediate range
of temperature, $\hbar\omega_c\ll T\ll
\hbar\omega_c\ln(\omega_c\tau_L)$, there is an exponentially
fast crossover between the two regimes. The regime most relevant to
the experiments \cite{zudov01,mani02,zudov03,dorozhkin03,willett03} is
that of high temperature, $T\agt\hbar\omega_c$. It is worth noting
that Eqs.~(\ref{21a}),(\ref{21}) remain valid in the regime of strong
heating as well, provided the effective electron temperature $T_e$
[Eq.~(\ref{6})] is substituted for $T$.

For convenience, we also
reproduce here $\Delta\sigma_{\rm ph}^{(q)}$ in the case of
overlapping Landau levels; specifically, for $|\omega\pm\omega_c|\agt
\omega_c$ in the regime linear with respect to the microwave power
(see Eq.~(8) of
Ref.~\onlinecite{dmitriev03a}; here we omit numerical factors):
\begin{eqnarray}
\Delta\sigma_{\rm ph}^{(q)}&\sim& -\sigma_0^{\rm D}\,{\tau_{\rm
 ee}\over\tau}\,\left({e{\cal E}_\omega v_F\over
\hbar\omega^2}\right)^2{\omega\over\omega_c} \nonumber\\
&\times&\sin{2\pi\omega\over\omega_c}\,\exp \left(-{2\pi\over
\omega_c\tau_q}\right)~. \label{22}
\end{eqnarray}
The electron-electron scattering time \cite{dmitriev03a} $\tau_{\rm
ee}\propto
T_e^{-2}$ (up to a logarithmic factor) depends on the effective electron
temperature $T_e$ [Eq.~(\ref{6})].
Although both contributions, Eqs.~(\ref{21}),(\ref{22}) have the same
period in $\omega/\omega_c$, crucial distinctions are clear.

Firstly, their phases are shifted by $\pi/2$. Secondly, despite both
contributions being proportional to a certain inelastic relaxation
time, they are different in that the amplitude of $\Delta\sigma_{\rm
ph}^{(q)}$ is limited by $\tau_{\rm ee}$ (which at low $T$ is much
shorter than the electron-phonon scattering time), whereas the
classical term is not sensitive to the inelastic electron-electron
scattering in any essential way \cite{ee} and is proportional to the
energy relaxation time [$\tau_{\rm in}$ in Eqs.~(\ref{7}),(\ref{21})],
limited by coupling to the external bath (phonons). It follows that in
the limit of small $T$ the ratio of the amplitudes of the OPC,
classical-to-quantum, contains a large $T$ dependent factor $\tau_{\rm
in}/\tau_{\rm ee}$, which may be easily as large as $10^2$. The
sensitivity of $\Delta\sigma_{\rm ph}^{(q)}$ to electron-electron
collisions stems from the fact that the quantum contribution is due to
a radiation-induced change of the distribution function
$f(\varepsilon)$ that oscillates with {\it both} $\varepsilon$ and
$\omega$. By contrast, the classical contribution $\Delta\sigma_{\rm
ph}^{(c)}$ is associated with an oscillatory term in the
characteristic electron temperature, i.e., with a smooth part of
$f(\varepsilon)$ which oscillates with $\omega$
only. \cite{quantum_heating}

Thirdly, the dependences of the envelope of the OPC on $\omega$,
$\omega_c$, and the degree of disorder are quite different. The most
important point is that although there is a small factor $\propto
\epsilon_F^{-1}$ in Eq.~(\ref{21}), in addition to another small
factor $a/\delta$, the {\it damping} of the classical term is much
weaker than that of $\Delta\sigma_{\rm ph}^{(q)}\propto \exp
(-2\pi/\omega_c\tau_q)$. Indeed, the exponential damping of
$\Delta\sigma_{\rm ph}^{(c)}$ is governed by $\tau_L$
[Eq.~(\ref{21})], which is far larger than $\tau_q$ in high-mobility
samples. It is only that in the limit of very low $B$ that the
$\omega_c^{-3}$ factor in the exponent of Eq.~(\ref{21}) suppresses
the classical OPC more effectively than the linear in $\omega_c^{-1}$
Dingle factor in the quantum case.

It is important to stress that the amplitude of the classical OPC in
units of the dark conductivity is not large under the conditions
of the experiments on the zero-resistance states. Indeed, the
pre-exponential factor of Eq.~(\ref{21}) may be written as
$\sigma_0^{\rm D}(\Delta_h^2/\epsilon_F T)(a/\delta)$ for the regime
linear with respect to the microwave power. Now, the crossover to the
regime of strong heating occurs when the classical OPC is still small,
namely the ratio $\Delta\sigma_{\rm ph}^{(c)}/\sigma_0^{\rm D}$ is of
order $(T/\epsilon_F)(a/\delta)\alt 10^{-2}$. This should be
contrasted with the quantum OPC which may become large (and thus lead
to the zero-resistance states) when the heating may be still
negligible. For the regime of strong heating, when the effective
electron temperature $T_e\agt T$, the amplitude of the
classical OPC shows a sub-linear growth with increasing microwave
power and may be estimated as $\sigma_0^{\rm
D}(T_e/\epsilon_F)(a/\delta)$. In particular, for the piezoelectric
mechanism of the energy relaxation due to electron-phonon coupling,
the classical OPC grows as ${\cal E}_\omega^{2/5}$ [see the discussion
below Eq.~(\ref{10})]. We conclude that, because of the slow growth
with increasing microwave power, the characteristic ratio
$\Delta\sigma_{\rm ph}^{(c)}/\sigma_0^{\rm D}$ can hardly exceed the
level of a few percent in the current experiments. That is to say the
zero-resistance states are related to the {\it quantum} OPC. The most
favorable conditions for the observation of the classical OPC should
be realized if the quantum contribution to the OPC is strongly damped,
which means sufficiently large $2\pi/\omega\tau_q\agt 7$. For a
high-mobility sample with $\tau_q\sim 3\,$ps, this would require
$\omega/2\pi\alt 50\,$GHz.

The above analysis shows that the classical OPC cannot possibly explain the
experimentally reported strong deviations of the ratio $\tau_{\rm ph}/\tau_q$
from 1/2, the value predicted by the theory of the quantum OPC, as discussed
in Introduction. We thus argue that the experiments on the damping of
Shubnikov-de Haas oscillations might strongly overestimate the single-particle
scattering rate $\tau_q^{-1}$. One of the reasons could be the presence of
macroscopic inhomogeneities leading to an inhomogeneous broadening of Landau
levels, which might be by far larger than the homogeneous broadening given by
$\tau_q^{-1}$ and measured in the photoconductivity experiments (such a
possibility was mentioned in Ref.~\onlinecite{zudov01}). We suggest that
measuring the damping of the OPC provides a reliable means of extracting
$\tau_q^{-1}$ from the magnetooscillations, free from the effect of the
additional inhomogeneous damping characteristic to the Shubnikov-de Haas
measurements. The method based on the OPC is particularly useful in
high-mobility samples, where $\tau_q^{-1}$ is small and in the conventional
Shubnikov--de Haas measurements one has to go to fairly low temperatures to
separate the impurity-induced damping from that related to the thermal
smearing of the Fermi surface.

\section{Conclusions}

In summary, we have analyzed the quasiclassical mechanism of
magnetooscillations in the ac- and photoconductivity, related to non-Markovian
dynamics of disorder-induced scattering of electrons in high-mobility
structures. We have calculated the leading contribution associated with a
radiation-induced change of the electron distribution function, which is
proportional to the inelastic (electron-phonon) relaxation time. We have found
that the quasiclassical oscillations in the photoconductivity are weak under
the conditions of current experiments. Therefore, the zero-resistance states
and the strong oscillations that have been observed in the experiments are
likely due to the quantum mechanism of
Refs.~\onlinecite{dmitriev03,dmitriev03a}. We argue that the damping of the
oscillatory photoconductivity provides a reliable method of measuring the
homogeneous broadening of Landau levels (single-particle scattering rate
$\tau_q^{-1}$ ) in high-mobility structures (which also resolves the dilemma
posed in Introduction: the analysis of the damping of Shubnikov-de Haas
oscillations apparently gives overestimated values of $\tau_q^{-1}$ due to
an inhomogeneous broadening).

On the other hand, we have identified a range of parameters within which the
quasiclassical mechanism yields oscillations of the photoconductivity that may
dominate at {\it small} $B$ over those based on the Landau quantization. In
addition to the different low-$B$ damping factor, the quasiclassical
oscillations are shifted in phase by $\pi/2$ with respect to the quantum
oscillations, see Eqs.~(\ref{21}),(\ref{22}).  We have also shown that the
quasiclassical magnetooscillations in the {\it ac conductivity} are much
stronger than in the photoconductivity and may easily compete with the quantum
oscillations. \cite{dmitriev03}

We thank R.~R.~Du, K.~von~Klitzing, R.~G.~Mani, J.~H.~Smet, and M.~A.~Zudov
for information about the experiments. We are grateful to I.~L.~Aleiner and
I.~V.~Gornyi for valuable discussions. In particlular, we thank I.~L.~Aleiner
for attracting our attention to the importance of quantum interaction
corrections. This work was supported by the SPP ``Quanten-Hall-Systeme" of DFG
and by RFBR.

\appendix*

\section{Return probability in a magnetic field}

The return probability
$P_\omega$ [Eq.~(\ref{14})] can be directly
evaluated by using the quasiclassical propagator $D=(L_0+\tau_S^{-1})^{-1}$ 
[Eq.~(\ref{D})].  
In this Appendix, we present a different, more illustrative way to derive $P_\omega$. We recall that
the Liouville operator $L_0$ [Eq.~(\ref{12})]
represents the time evolution
of the direction of the electron velocity ${\bf n}={\bf v}/|{\bf v}|=(-\sin\phi,\,\cos\phi)$ 
as a combination of the cyclotron motion and the angle diffusion due to
scattering off smooth disorder. The random part $\chi$ of the angle $\phi$ is characterized by 
a white noise spectrum of $\partial_t\chi$:
\bea
\nonumber
&&\phi(t)=\phi_0+\omega_c t+\chi(t)\,,\\
&&\langle\,\partial_t\chi(t)\,\partial_{t^\prime}\chi(t^\prime)\,\rangle=
\frac{2}{\tau_L}\delta(t-t^\prime).
\label{wn}
\eea
In what follows we calculate the mean-square fluctuation of the guiding
center of cyclotron motion $\delta$, Eq.~(\ref{delta}), and
mean-square shifts of 
an electron along and across the cyclotron orbit after $n$ cyclotron periods at  
$t={n\,T_c}=2\pi\,n/\omega_c$. 
For definiteness, let  
the guiding
center be initially placed at the origin, ${\bf R}(t=0)=(0,0)$, and 
the electron coordinate and velocity be  ${\bf r}\,(t=0)=(R_c,\,0)$,
${\bf v}(t=0)=(0,\,v_F)$ ($R_c=v_F/\omega_c$ is the cyclotron radius).   
Using Eq.~(\ref{wn}) we get the mean-square shifts of the position of
 guiding center ${\bf R}$ and the
fluctuating angle $\chi$ in time $t=nT_c$: 
\bea\nonumber
\label{shift1}
\langle\, R_x^2\,\rangle&=&
\left
\langle 
\left
( \int^{n\,T_c}_0\!\!\!
{\rm d}t\, R_c\,\cos\phi(t)\,\partial_t\chi(t)
\right
)^2
\right
\rangle\\\nonumber
&=&\frac{2\,R_c^2}{\tau_L}
\,\int^{n\,T_c}_0\!\!\!{\rm d}t\cos^2\!\phi(t)
=R_c^2\, {n\,T_c\over\tau_L}\\ \nonumber \phantom{a} \\\nonumber
\langle\, R_y^2\,\rangle&=&\langle\, R_x^2\,\rangle\,,\\ 
\langle\, \chi^2\,\rangle&=&
\left
\langle 
\left
( \int^{n\,T_c}_0\!\!\!
{\rm d}t \,\partial_t\chi(t)
\,\right)^2
\right
\rangle=\frac{2\,n\,T_c}{\tau_L}\,.
\eea
The diffusion approximation is valid as long as the root-mean-square shift
of the guiding center after one cyclotron revolution $(n=1)$ exceeds the characteristic length scale
of the random potential,
\be
\label{deltaApp}
\delta=\left[\,\langle\, R_x^2\,\rangle+\langle\, R_y^2\,\rangle\,\right]^{1/2}
=R_c\left(\frac{4\pi}{\omega_c\tau_L}\right)^{1/2}\gg d.
\ee
In the same manner, we calculate the 
mean square of electron shifts along and across
the cyclotron orbit, 
$x_\parallel\equiv y(t=n\,T_c)=\int^{n\,T_c}_0{\rm d}t\,v_F\,\cos\phi(t)$ and
$x_\perp\equiv x(t=n\,T_c)-R_c=-\int^{n\,T_c}_0{\rm d}t\,v_F\,\sin\phi(t)$, respectively,
\bea\nonumber
\label{shift2}
\langle x_\perp^2\rangle&=&
\left
\langle 
\left
( \int^{n\,T_c}_0\!\!\!
{\rm d}t\, v_F\,\cos(\omega_c t)\int^t_0\!{\rm d}t^\prime
\,\partial_{t^\prime}\chi(t^\prime)
\right
)^2
\right
\rangle\\\nonumber&=&\langle R_x^2\rangle\,, \\\nonumber\phantom{a}\\
\langle x_\parallel^2\rangle&=&\langle (R_y+R_c\,\chi)^2\rangle=
\langle R_y^2\rangle+R_c^2\,\langle\chi^2\rangle\,.
\eea
It follows that fluctuations along the cyclotron orbit
are enhanced with respect to those across the orbit,
 $\langle x_\parallel^2\rangle=3\langle x_\perp^2\rangle=3\,n\,\delta^2/2$, and
we arrive at the anisotropic electron distribution after $n$ cyclotron
revolutions,
\bea\label{p_1}
p_n(x_\perp,x_\parallel)
&=&\frac{1}{\sqrt{\pi}\,n\,\delta}\exp\left(- \frac{x_\perp^2}{n\,\delta^2}\right)
\nonumber \\ &\times&\frac{1}{\sqrt{3\pi}\,n\,\delta}\exp\left(- \frac{x_\parallel^2}{3\,n\,\delta^2}\right),
\eea
which enters Eq.~(\ref{15}) for the return probability.


\begin{thebibliography}{1}

\bibitem[*]{byline} Also at A.F.~Ioffe Physico-Technical
Institute, 194021 St.~Petersburg, Russia.

\bibitem[$\dagger$]{} Also at Petersburg Nuclear Physics
Institute, 188350 St.~Petersburg, Russia.

\bibitem{zudov01} M.A.~Zudov, R.R.~Du, J.A.~Simmons, and J.L.~Reno,
Phys. Rev. B {\bf 64}, 201311(R) (2001); P.D.~Ye, L.W.~Engel,
D.C.~Tsui, J.A.~Simmons, J.R.~Wendt, G.A.~Vawter, and J.L.~Reno,
Appl.\ Phys.\ Lett.\ {\bf 79}, 2193 (2001).

\bibitem{mani02} R.G.~Mani, J.H.~Smet, K.~von~Klitzing, V.~Narayanamurti,
W.B.~Johnson, and V.~Umansky, Nature {\bf 420}, 646 (2002); Phys.\ Rev.\ B
{\bf 69}, 193304 (2004); cond-mat/0306388.

\bibitem{zudov03} M.A.~Zudov, R.R.~Du, L.N.~Pfeiffer, and K.W.~West,
Phys. Rev. Lett. {\bf 90}, 046807 (2003); C.L.~Yang, M.A. Zudov,
T.A. Knuuttila, R.R. Du, L.N. Pfeiffer, and K.W. West, {\it ibid.}
{\bf 91}, 096803 (2003).

\bibitem{dorozhkin03} S.I. Dorozhkin, JETP Lett.\ {\bf 77}, 577
(2003).

\bibitem{willett03} R.L.~Willett, L.N.~Pfeiffer, and K.W.~West,
Phys.\ Rev.\ Lett.\ {\bf 93}, 026804 (2004).

\bibitem{andreev03} A.V.~Andreev, I.L.~Aleiner, and A.J.~Millis,
Phys.\ Rev.\ Lett.\ {\bf 91}, 056803 (2003).

\bibitem{dmitriev03} I.A.~Dmitriev, A.D.~Mirlin, and D.G.~Polyakov,
Phys.\ Rev.\ Lett. {\bf 91}, 226802 (2003).

\bibitem{dmitriev03a} I.A.~Dmitriev, M.G.~Vavilov, I.L.~Aleiner,
A.D.~Mirlin, and D.G.~Polyakov, cond-mat/0310668.

\bibitem{durst03} A.C.~Durst, S.~Sachdev, N.~Read, and S.M.~Girvin,
Phys.\ Rev.\ Lett.\ {\bf 91}, 086803 (2003).

\bibitem{ryzhii} V.I.~Ryzhii, Sov.\ Phys.\ Solid State {\bf 11}, 2078
(1970); V.I.~Ryzhii, R.A.Suris, and B.S.~Shchamkhalova, Sov.\ Phys.\
Semicond.\ {\bf 20}, 1299 (1986).

\bibitem{vavilov03} M.G.~Vavilov and I.L.~Aleiner, Phys.\ Rev.\ B {\bf 69},
035303 (2004).

\bibitem{sdh} To avoid confusion, it is worthwhile to note that the
Shubnikov-de Haas oscillations of the dark resistivity were damped in
Refs.~\onlinecite{zudov01,mani02,zudov03,dorozhkin03,willett03} mostly by
temperature rather than by disorder.

\bibitem{mem} For smooth disorder, manifestations of the memory effects in the
magnetoresistance and {\it ac} transport were studied in A.D.~Mirlin,
J.~Wilke, F.~Evers, D.G.~Polyakov, and P.~W\"olfle, Phys.\ Rev.\ Lett.\ {\bf
83}, 2801 (1999); J.~Wilke, A.D.~Mirlin, D.G.~Polyakov, F.~Evers, and
P.~W\"olfle, Phys.\ Rev.\ B {\bf 61}, 13774 (2000). Recently, the role of the
memory effects in the photoconductivity, in the case of smooth disorder, was
discussed in terms of the influence of microwave radiation on the collision
integral in Ref.~\onlinecite{vavilov03} and in I.L.~Aleiner, B.L.~Altshuler,
and A.V.~Andreev, unpublished. However, the memory effects manifest themselves
in the photoconductivity much more strongly through a radiation-induced change
of the electron distribution function.

\bibitem{mirlin01} A.D.~Mirlin, D.G.~Polyakov, F.~Evers, and
P.~W\"olfle, Phys.\ Rev.\ Lett.\ {\bf 87}, 126805 (2001).

\bibitem{umansky} V.~Umansky, R.~de Picciotto, and M.~Heiblum,
Appl.\ Phys.\ Lett.\ {\bf 71}, 683 (1997). 

\bibitem{zala01} G.~Zala, B.N.~Narozhny, and I.L.~Aleiner, Phys.\
Rev.\ B {\bf 64}, 214204 (2001).

\bibitem{gornyi03} I.V.~Gornyi and A.D.~Mirlin, Phys.\ Rev.\ B {\bf 69},
045313 (2004).

\bibitem{ee} More precisely, the effect of inelastic electron-electron
scattering on the OPC is twofold: firstly, it suppresses oscillations
of $f(\varepsilon)$ induced by those of the DOS; secondly, it affects
the exact shape of the smooth part of $f(\varepsilon)$ by thermalizing
electrons among themselves. The former effect leads to the damping of
the quantum term $\Delta\sigma_{\rm ph}^{(q)}$, as discussed in
Ref.~\onlinecite{dmitriev03a}. The latter can only yield a numerical
coefficient of order unity in the classical term $\Delta\sigma_{\rm
ph}^{(c)}$, depending on what other mechanisms of inelastic scattering
are. Moreover, this numerical factor is one, so that electron-electron
inelastic scattering does not manifest itself in $\Delta\sigma_{\rm
ph}^{(c)}$ at all, if the inelastic coupling to the thermal bath in
the presence of a driving force yields $f(\varepsilon)$ whose shape is
given by the Fermi distribution (but with an effective temperature
different from that of the bath). This is the case, e.g., for the
Fokker-Planck mechanism of Eqs.~(\ref{9}),(\ref{10}).

\bibitem{polyakov02} Similar oscillations of the {\it ac} conductivity
were studied for the case of a random antidot array, where the memory
effects are still stronger, in D.G.~Polyakov, F.~Evers, and
I.V.~Gornyi, Phys. Rev. B {\bf 65}, 125326 (2002). The antidot-array
model also describes correctly rare strong short-ranged scatterers in
the absence of background smooth disorder.

\bibitem{rudin97} Equation (\ref{18a}) has the same structure as an
interaction-induced correction to the tunneling DOS, studied in
A.M.~Rudin, I.L.~Aleiner, and L.I.~Glazman, Phys.\ Rev.\ Lett.\ {\bf
78}, 709 (1997), which is also proportional to the return probability
$P_\omega$.

\bibitem{quantum_heating} It is worth noting that there exists also a
subleading quantum contribution to the OPC that is due to the
oscillatory heating, i.e., due to a smooth part of $f(\varepsilon)$
which oscillates with $\omega$ (because of the oscillatory absorption
rate coming in turn from the Landau quantization). It has the same
damping factor as in Eq.~(\ref{22}), but the phase of the oscillations
as in Eq.~(\ref{21}). This quantum term is, however, much smaller than
given by Eq.~(\ref{22}); specifically, by a factor
$(\hbar^2\omega\omega_c/\epsilon_F^2)\tau_{\rm in}/\tau_{\rm ee}\ll
1$.


\end{thebibliography}
\end{document}